\documentclass[12pt]{article}
\usepackage{amssymb}

\def\eps{\epsilon}
\def\l{\lambda}
\def\be{\begin{equation}}
\def\ee{\end{equation}}
\def\ba{\begin{eqnarray}}
\def\ea{\end{eqnarray}}
\newcommand{\nn}{\nonumber}
\newcommand{\no}{\nonumber \\}

\def\S{{\cal S}}

\def\l{\lambda}
\def\b#1{{\mathbb #1}}

\newcommand{\T}{\mbox{Tr}}
\begin{document}
\begin{titlepage}
September 2000         \hfill
\vskip -0.55cm 
\hfill    UCB-PTH-00/27  
 
\hfill  LBNL-46775  
\begin{center}

\vskip .15in

\renewcommand{\thefootnote}{\fnsymbol{footnote}}
{\large \bf Properties of perturbative solutions of unilateral matrix 
equations}
\vskip .25in
Bianca L. Cerchiai\footnote{email address: BLCerchiai@lbl.gov} and 
Bruno Zumino\footnote{email address: zumino@thsrv.lbl.gov}
\vskip .25in

{\em    Department of Physics  \\
        University of California   \\
                                and     \\
        Theoretical Physics Group   \\
        Lawrence Berkeley National Laboratory  \\
        University of California   \\
        Berkeley, California 94720}
\end{center}
\vskip .25in

\begin{abstract}

A left-unilateral matrix equation is an algebraic equation of the form
$$
a_0+a_1 x+a_2 x^2+\ldots +a_n x^n=0
$$
where the coefficients $a_r$ and the unknown $x$ are square matrices of the
same order and all coefficients are on the left 
(similarly for a right-unilateral equation). Recently certain perturbative
solutions of unilateral equations and their properties have been
discussed. We present a unified approach based on the generalized Bezout 
theorem for matrix polynomials. Two equations discussed in the literature,
their perturbative solutions and the relation between them are described.
More abstractly, the coefficients and the unknown can be taken as elements of
an associative, but possibly noncommutative, algebra.
\end{abstract}
\end{titlepage}

\newpage

\section{Introduction}
\setcounter{footnote}{0}
In a discussion of generalized Born-Infeld theories~\cite{BMZ, ABMZ1} 
the construction of the Lagrangian was reduced to the solution of certain
unilateral algebraic matrix equations and it was conjectured that the 
iterative solution of those equations is a sum of symmetric 
polynomials in the coefficients and  of terms which are commutators.
Equivalently, that the trace of the matrix solution is equal to the sum of
traces of symmetric polynomials in the coefficients. The conjecture was later
proven in \cite{ABMZ2} and by A. Schwarz in \cite{Schwarz} using
different methods and a slightly different form of the equation.

In the present note we combine Schwarz' s idea of expressing the trace of the
solution as a contour integral in the complex plane of the trace of the
resolvent of the corresponding matrix with the idea used in 
\cite{ABMZ2} of using the basic property of the trace of the logarithm
of matrices.

The two approaches can be easily combined by using the generalized Bezout
theorem for matrix polynomials. The qualitative algebraic fact that the 
trace of the solution is given by a sum of traces of symmetrized polynomials
in the coefficients emerges almost without any computation. The coefficients
in the expansion are not hard to compute and we give an explicit expression for
them.
\footnote{Our contour integral formulas are very similar to Schwarz' s, but
there seem to be some minor discrepancies between ours and his.}
The method used here applies equally to the two different equations considered
in \cite{ABMZ2} and \cite{Schwarz}; in the last section we clarify the
relation between the two equations and their solutions.

Before closing this section we recall the statement of the generalized Bezout
theorem (see e.g. \cite{Gant}). Let
\be
P(\l)=a_0+a_1 \l+a_2 \l^2+\ldots+a_n \l^n, \qquad \l \in \b{C}
\label{characteristic}
\ee
be a polynomial with square matrix coefficients $a_r$ and $x$ a square matrix
of the same order. Define
\be
P(x)=a_0+a_1 x+a_2 x^2+\ldots+ a_n x^n
\ee
with the coefficients all on the left. It is easy to verify that
\be
P(\l)-P(x)=Q(\lambda,x)(\lambda-x)\: ,
\label{bezout1}
\ee
where
\be
Q(\lambda,x)=\sum_{l=0}^{n-1}\lambda^l \left(\sum_{r=l+1}^n \eps a_r 
x^{r-l-1}\right)\: .
\ee
In other words $\l-x$ is a divisor of $P(\l)-P(x)$ on the right (if we had
taken all the coefficients on the right, then $\l-x$ would have been
a divisor on the left).

We shall study matrix equations of the type
\be
P(x)=0\: ,
\label{equation}
\ee
which we shall call left-unilateral matrix equations (meaning that all the
coefficients are on the left). If $x$ is a solution of (\ref{equation}),
the characteristic polynomial $P(\l)$ of the equation can be factorized as
\be
P(\l)=Q(\lambda,x)(\lambda-x)\: .
\ee
It is clear that the Bezout theorem applies more abstractly if one considers
$a_r$ and $x$ as elements of an associative, but possibly noncommutative,
algebra. The same remark applies to the rest of this note, if one uses
an appropriate algebraic definition of the trace as cyclic average
(see \cite{ABMZ1}).

\section{The Trace of the Solutions of Unilateral Matrix Equations}
\setcounter{equation}{0}
A. Schwarz~\cite{Schwarz} considers the unilateral matrix equation
\be
x^n=1+\epsilon\left(a_0+a_1 x+\ldots+a_{n-1} x^{n-1} \right)\: .
\label{schwarz1}
\ee
where $\eps$ is a small parameter. For $\eps=0$ the equation has $n$ solutions
in the complex plane, the $n$ roots of $1$. For $\eps \neq 0$ each of these
solutions admits a perturbative expansion as a formal power series in
$\eps$. A. Schwarz proves that $\T \; x^s$ 
can be expressed as a power series whose coefficients are symmetrized products
of the $a_i$ and gives an expression for the coefficients 
in terms of contour integrals in the complex plane.

We have found that a more explicit expression for $\T \: x^s$ is provided by
\be
\T \: x^s =\T \: 1+s \sum_{k=1}^{\infty} \frac{\eps^k}{n^k} 
\! \! \sum_{n_0+ \ldots +n_{n-1}=k} \! \!
\frac{\T {\cal S} (a_0^{n_0} \ldots a_{n-1}^{n_{n-1}})}{n_0!\ldots n_{n-1}!} 
\prod_{r=1}^{k-1} \left(s+\sum_{l=1}^{n-1} l n_l-rn \right)\: .
\label{sol1}
\ee
This formula holds for positive as well as for negative values of $s$.
Here the expansion is made around $1$. A similar expansion
could be derived for all the $n$ iterative solutions of the equation.
The normalization of the symmetrized product 
$\S(a_0^{n_0} \ldots a_n^{n_n})$
is chosen in such a way as to give the 
ordinary product if the factors commute \cite{ABMZ2}.

If we introduce the notation
\be
\left(
\begin{array}{l}
\alpha \\ k
\end{array}
\right)
=\left\{ 
\begin{array}{ll}
\displaystyle{\prod_{r=1}^k} \frac{\alpha-r+1}{r} & \mbox{ for } k=1,2,\ldots 
\\
\\
1 & \mbox{ for } k=0\: ,
\end{array}
\right.
\ee
which reduces to the usual definition 
of $\left( \begin{array}{l} \alpha \\ k \end{array} \right)$ 
for $\alpha \in \b{N}$, then the result~(\ref{sol1}) can be rewritten
\ba
\lefteqn{\T \: x^s=}
\label{sol2}\\
&& \! \! \! \! \! \! \! \! \!
\T \: 1+\frac{s}{n} \sum_{k=1}^{\infty} \eps^k (k-1)! 
\! \! \! \! \! \sum_{n_0+ \ldots +n_{n-1}=k} \! \! \! \! \! \!\! \!
\frac{\T {\cal S} (a_0^{n_0} \ldots a_{n-1}^{n_{n-1}})}{n_0!\ldots n_{n-1}!} 
\left(
\begin{array}{l}
\frac{1}{n}(s-n+\sum_{l=1}^{n-1} l n_l)\\ \\
k-1
\end{array}
\right). \nn
\ea

As a first step to prove (\ref{sol1}), we apply the generalized Bezout 
theorem. For (\ref{schwarz1}) the characteristic polynomial is
\be 
P(\l) \equiv 1-\l^n+\epsilon\left(a_0+a_1 \l+\ldots +a_{n-1} \l^{n-1} \right)
\ee
and, if $P(x)=0$,
\be
P(\lambda)=Q(\lambda,x)(\lambda-x)
\label{bezout2}
\ee
with
\be
Q(\lambda,x)=\sum_{l=0}^{n-1} \left(\sum_{r=l+1}^{n-1} \eps a_r 
x^{r-l-1}-x^{n-l-1} \right) \lambda^l \: .
\ee
Let $x$ be one of the solutions of (\ref{schwarz1}), e.g. the one which
reduces to $1$ for vanishing $\eps$. (For the other solutions a similar
procedure can be followed).
We take the logarithm and then the trace of (\ref{bezout2}), and obtain
\be
\T \log P(\lambda)=\T \log(Q(\lambda,x))+\T \log(\lambda-x) \: .
\ee
Differentiation with respect to $\lambda$ yields
\be
\T \frac{1}{P(\l)}P'(\l)=\T \frac{1}{Q(\l,x)}Q'(\l,x)+ \T \frac{1}{\l-x}\: .
\ee
More generally, we can multiply the above equation by any function $f(\l)$ 
of $\l$, which is regular in a neighbourhood of $1$, so as to get
\be
\T \frac{1}{P(\l)}P'(\l) f(\l)=\T \frac{1}{Q(\l,x)}Q'(\l,x) f(\l)+
\T \frac{1}{\l-x} f(\l)\: .
\label{argument}
\ee

As a next step, we follow Schwarz' s idea of using a contour integration 
to isolate the relevant part in the trace.
We consider a small circle $\Gamma$ around $1$ in the complex plane, or more
generally a small closed curve winding once around~$1$, and containing no other
solution of (\ref{schwarz1}) for $\eps=0$. Then we compute the 
integral of equation (\ref{argument}) along it for small $\eps$
\ba
\lefteqn{(2 \pi i)^{-1} \oint_{\Gamma} d \l \: 
\T \frac{1}{P(\l)}P'(\l) f(\l)=}\\
&&(2 \pi i)^{-1} \oint_{\Gamma} d \l \: \T \frac{1}{Q(\l,x)}Q'(\l,x) f(\l)
+(2 \pi i)^{-1} \oint_{\Gamma} d \l \: \T \frac{1}{\l-x} f(\l)\: . \nn
\ea
The two integrals on the right hand-side can be evaluated through the Cauchy
theorem in the following way. We are considering the case of small $\eps$,
and then $Q$ is a polynomial in $x$ and $\l$, which vanishes for $\l$ near 
$e^{\frac{2 \pi i k}{n}}$, $k=1,\ldots,n-1$, so that in this case 
its inverse $Q^{-1}$ has no singularities near $1$. 
$Q'$ and $f$ are regular functions near $1$ and
therefore the Cauchy theorem guarantees that the first
term on the right hand-side vanishes:
\be
(2 \pi i)^{-1} \oint_{\Gamma} d \l \: \T \frac{1}{Q(\l,x)}Q'(\l,x) f(\l)=0\: .
\ee
On the other hand $\T \frac{1}{\l-x} f(\l)$ has poles
for $\l$ near $1$ with total residue $\T f(x)$. As $x$ is close to $1$ 
for small $\eps$, the Cauchy theorem yields
\be
(2 \pi i)^{-1} \oint_{\Gamma} d \l \: \T \frac{1}{\l-x} f(\l)=\T f(x)\: .
\ee
Finally, we obtain
\be
\T f(x)=(2 \pi i)^{-1} \oint_{\Gamma} d \l \: \T \frac{1}{P(\l)}P'(\l)
f(\l)\: .
\label{integral}
\ee
Therefore, the problem of computing $\T f(x)$ amounts to evaluating the
integral on the right hand-side of (\ref{integral}).

We can factorize
\be
P(\l)=(1-\l^n) T(\l) \: ,
\label{factor}
\ee
where
\be
T(\l)=1-\eps (\l^n-1)^{-1} \sum_{l=0}^{n-1} a_l \l^l \: .
\ee
Then
\be
\frac{1}{P(\l)}P'(\l)=-\frac{n \l^{n-1}}{1-\l^n}+\frac{1}{T(\l)}T'(\l)
\ee
and (\ref{integral}) becomes
\be
\T \: f(x)=\frac{1}{2 \pi i} \oint_{\Gamma} d \l \: 
\frac{n \l^{n-1}}{\l^n-1}\T \: f(\l)+\frac{1}{2 \pi i} \oint_{\Gamma} d \l \: 
\T \frac{1}{T(\l)} T'(\l) f(\l) \: .
\ee
The first integral on the right hand-side equals $\T \: f(1)$.
To evaluate the second term we integrate by parts
\be
\frac{1}{2 \pi i} \oint_{\Gamma} d \l \: \T \frac{1}{T(\l)} T'(\l) f(\l)=
-\frac{1}{2 \pi i} \oint_{\Gamma} d \l \: \T \log(T(\l)) f'(\l) \: .
\ee
This is justified, because $\T \log T(\l)$ and $f(\l)$ are regular
on $\Gamma$, so that
\be
\frac{1}{2 \pi i} \oint_{\Gamma} d \l \: \T \frac{d}{d\l}\left( 
\log(T(\l)) f(\l)\right)=0.
\ee
In this way we obtain
\be
\T \: f(x)=\T \: f(1) -\frac{1}{2 \pi i} \oint_{\Gamma} d \l \: 
\T \log(T(\l)) f'(\l) \: .
\label{partial}
\ee
We expand the logarithm
\be
-\log(1-\alpha)=\sum_{k=1}^{\infty} \frac{\alpha^k}{k} 
\: \: \mbox{ for $\alpha$ small}
\ee
and then make the following change of variable
\be
y=\l^n\: .
\label{transf}
\ee
This is possible, because if we start with a small closed curve $\Gamma$ 
winding once around $1$, then after the variable 
transformation (\ref{transf}) we still have a closed curve winding 
once around $1$. We restrict ourselves to the case $f(\l)=\l^s$.
\ba
\lefteqn{-\frac{s}{2 \pi i} \oint_{\Gamma} d \l \: \T \log\left(1-\eps 
\frac{\sum_{l=0}^{n-1} a_l \l^l}{\l^n-1}\right) \l^{s-1}=}
\label{steps} \\
&& s \sum_{k=1}^{\infty} \frac{\eps^k}{2 \pi i k} \oint_{\Gamma} d \l \T  
\left(\frac{\sum_{l=0}^{n-1} a_l \l^l}{\l^n-1}\right)^k \l^{s-1}= \no
&& \frac{s}{n} \sum_{k=1}^{\infty} \frac{\eps^k}{2 \pi i k} 
\oint_{\Gamma} dy \T  
\left(\frac{\sum_{l=0}^{n-1} a_l y^{\frac{l}{n}}}{y-1}\right)^k 
y^{\frac{s}{n}-1}\: . \nn
\ea
Notice that $y^{\frac{l}{n}}$ is regular in a neighbourhood of $1$.
In this form we can already see that the result is symmetrized in
the coefficients $\{a_i\}$, because
they enter through expressions of the type
\be
\left(\sum_{l=0}^{n-1} a_l y^{\frac{l}{n}} \right)^k=
\sum_{n_0+\ldots +n_{n-1}=k}
\frac{k!}{n_0 ! \ldots n_{n-1}!} \S(a_0^{n_0} \ldots a_{n-1}^{n_{n-1}})
y^{\displaystyle \frac{1}{n}\sum_{l=1}^{n-1} l n_l}\: .
\ee
We obtain 
\ba
\lefteqn{\T \: x^s}\\
&& \! \! \! \! \! \!\! \! \! \!   =\T \: 1+\frac{s}{n} \sum_{k=1}^{\infty} \eps^k 
\! \! \!  \sum_{n_0+\ldots +n_{n-1}=k} \! \! \! 
\frac{\T \S(a_0^{n_0} \ldots a_{n-1}^{n_{n-1}})}{n_0 ! \ldots n_{n-1}!}
\frac{d^{k-1}}{dy^{k-1}} 
y^{\displaystyle \frac{1}{n}(\sum_{l=1}^{n-1} l n_l+s-n)} 
\Big|_{y=1} \no
&& \! \! \! \! \! \!\! \! \! \! =\T \: 1 +s 
\sum_{k=1}^{\infty} \frac{\eps^k}{n^k} 
\sum_{n_0+ \ldots +n_{n-1}=k}
\frac{\T \S (a_0^{n_0} \ldots a_{n-1}^{n_{n-1}})}{n_0!\ldots n_{n-1}!} 
\prod_{r=1}^{k-1} \left(s+\sum_{l=1}^{n-1} l n_l-rn \right)\: .
\nn
\ea
Here we have applied the Cauchy theorem in its more general form
\be
(2 \pi i)^{-1} \oint_C dy \frac{f(y)}{(y-y_0)^k}
=\frac{1}{k-1!} \frac{d^{k-1}}{dy^{k-1}} f(y) |_{y=y_0}\: ,
\ee
where $C$ a closed curve winding once around $y_0$, and $f(y)$ is
a function which is regular inside $C$.

In the simplest case $n=2$ it is possible to solve the classical equation
(i.e. the equation for commuting $a_i$) and this provides a closed 
expression for (\ref{sol1})
\be
\T \: x=\T \left[ \eps \frac{a_1}{2} + \S
\sqrt{1+\eps a_0+(\eps \frac{a_1}{2})^2} \; \right]\: ,
\label{n2}
\ee
where we again have chosen the solution which reduces to $1$ for $\eps=0$.
In the appendix we show how expanding the square root in (\ref{n2}) near $1$ 
we actually recover (\ref{sol1}) in this case.

\section{Relation between two unilateral matrix \\ equations}
\setcounter{equation}{0}
The same procedure which we have used to prove (\ref{sol1}), 
namely applying the generalized Bezout theorem and then the Cauchy theorem,
can be equally well followed for the equation
\be
\Phi=A_0 +A_1 \Phi+ \ldots A_n \Phi^n\: ,
\label{ABMZ}
\ee
which is studied in \cite{ABMZ2}. If we define
\be
A(\l)=A_0 +A_1 \l+ \ldots A_n \l^n
\ee
the characteristic polynomial of (\ref{ABMZ}) is $\l-A(\l)$ and the result 
which corresponds to (\ref{partial}) is
\be
\T f(\Phi)=-\frac{1}{2 \pi i} \oint_C d \l \: \T \log(1-\frac{A(\l)}{\l}) 
f'(\l)\: .
\ee
Here $C$ is a closed curve winding once around $0$. We consider
the particular case $f(\l)=\l^s$ and expand the logarithm
\be
\T \: \Phi^s=s \sum_{k=1}^{\infty} \frac{(k-1)!}{2 \pi i} 
\sum_{n_0+\ldots n_n=k} 
\frac{\T \: \S(A_0^{n_0} \ldots A_n^{n_n})}{n_0! \ldots n_n !} \oint_C d \l \:
\l^{\sum_{l=0}^n (l-1)n_l+s-1}\: .
\ee
The Cauchy theorem has the effect of selecting the words of dimension 
$\sum_{l=0}^n (l-1)n_l=-s$ in the expansion of the logarithm, i.e. 
\be
\T \: \Phi^s=s \T \sum_{k=1}^{\infty} \frac{1}{k} (A_0+\ldots A_n)^k 
|_{\sum_{l=0}^n (l-1)n_l=-s}
\label{Dan}
\ee
and we recover the result already stated in \cite{ABMZ2}, where the
concept of dimension of a word was introduced.
It should be remarked that the solution to (\ref{ABMZ}) for vanishing
coefficients is $\Phi=0$. Consistent with that, as already noted in 
in  \cite{ABMZ2}, the result (\ref{Dan}) only holds for 
positive integers $s$: it has no sense to invert the solution in this case,
but see below.

To study more closely the relation between equation~(\ref{schwarz1}) 
and (\ref{ABMZ}) we make the Ansatz
\be
x=1+\alpha \Phi\: .
\label{ansatz}
\ee
(Again, the same procedure could be followed for any of the roots of unity
which are the solutions of (\ref{schwarz1}) for $\eps=0$.)

We choose the parameter $\alpha$ in such a way that
\be
\alpha^{n-1}=-n\: .
\label{choice}
\ee
There is no reason for $\alpha$ to be real, since most of the solutions
of (\ref{schwarz1}) are not real, even for real coefficients.

Then the relation between the $n$ coefficients $a_i$ and the $n+1$ 
coefficients $A_i$ is easily found to be
\be
A_l= \left\{
\begin{array}{ll}
 -\alpha^{l-n} \eps \displaystyle{\sum_{r=l}^{n-1}}
\left(\begin{array}{l} r\\ l \end{array} \right)
a_r & \mbox{ for } l=0,1 \\
\alpha^{l-n} \left(\begin{array}{l} n\\ l \end{array} \right)
-\alpha^{l-n} \eps \displaystyle{\sum_{r=l}^{n-1}}
\left(\begin{array}{l} r\\ l \end{array} \right)
a_r & \mbox{ for } 2 \le l \le n\: .
\end{array}
\right.
\label{transform}
\ee
Some remarks can be made with respect to (\ref{transform}).
First, observe that $A_n=1$ is fixed, but this had to be expected, because the
equation~(\ref{schwarz1}) has one coefficient less than the
equation~(\ref{ABMZ}).

Moreover, (\ref{transform}) is a linear relation, and it is invertible, 
so that once the result that the trace of the powers of a solution depends 
only on the symmetrized products of the coefficients is proven for one of
the two equations, it immediately follows also for the other.
Negative powers of $x$ can be expanded from (\ref{ansatz}) into a series of
positive powers of $\Phi$.

If $\eps$ is small, only the first two coefficients $A_0$ and 
$A_1$ are automatically small, the other coefficients satisfy 
\be
A_l \stackrel{\eps \rightarrow 0}{\rightarrow}
\alpha^{l-n} \left(\begin{array}{l} n\\ l \end{array} \right) \: \:
\mbox{ for }   2 \le l \le n\: .
\ee 
Therefore, the two expansions (\ref{sol1}) and (\ref{Dan}) do not 
necessarily hold for the same range of the coefficients.

\section*{Appendix}
\setcounter{section}{1}
\def\thesection{\Alph{section}}
\setcounter{equation}{0}

In this appendix we check that (\ref{sol1}) actually is the series 
expansion of (\ref{n2}) for $s=1$, $n=2$. We start by expanding the square 
root in (\ref{n2}) around $1$
\ba
\T \: x&=& \!\!\!\! \T \left[ 1+\eps \frac{a_1}{2}
+\sum_{r=1}^{\infty} (-1)^{r-1} \frac{1}{2^r r!} 
(2r-3) !! \S \left( \eps a_0+(\eps \frac{a_1}{2})^2 \right)^r \right] \no
&=& \!\!\!\!  \T \: 1+\eps \frac{\T \: a_1}{2} \\
&&\!\!\!\! +\sum_{r=1}^{\infty} (-1)^{r-1} 
\frac{1}{2^r} (2r-3) !! 
\sum_{n_0=0}^r \frac{\eps^{2r-n_0}}{n_0! (r-n_0)!}
\T\: \S\left(a_0^{n_0}\left(\frac{a_1}{2}\right)^{2(r-n_0)}\right)\: . \nn
\ea
We introduce the new variables:
\be
n_1=2(r-n_0), \quad k=n_0+n_1=2r-n_0\: .
\ee
Then we can rewrite
\ba
\T \: x&=& \! \! \!  \T \:1+\eps \frac{\T a_1}{2} \label{n=2}\\
&& \! \! \! +\sum_{k=1}^\infty \frac{\eps^k}{2^k} 
\sum_{\stackrel{\scriptstyle n_0+n_1=k}{n_1 even}} (-1)^{\frac{n_1}{2}+n_0-1}
\frac{(n_1-1)!!}{n_0! n_1!} (2 n_0+n_1-3)!! \T \S(a_0^{n_0} a_1^{n_1})\: ,
\nn
\ea
where we have used the relation
\be
\frac{n_1!}{\frac{n_1}{2}! 2^{\frac{n_1}{2}}}=(n_1-1)!! \: \: 
\mbox{ for $n_1$ even.}
\ee
Due to the conditions $n_0+n_1=k$, $n_1$ even, the following relation holds
\be
(-1)^{\frac{n_1}{2}+n_0-1} (n_1-1)!! (2 n_0+n_1-3)!!=
\prod_{r=1}^{k-1} \left(1+n_1-2r \right)
\ee
and so formula (\ref{n=2}) can be brought into the more compact form
\be
\T \: x=\T \: 1+\sum_{k=1}^\infty \frac{\eps^k}{2^k} \sum_{n_0+n_1=k}
\frac{1}{n_0! n_1!} \T \: \S(a_0^{n_0} a_1^{n_1})
\prod_{r=1}^{k-1} \left(1+n_1-2r \right)\: ,
\ee
which now coincides with (\ref{sol1}) for $s=1$, $n=2$.
It is no longer necessary to explicitly sum only over even values of $n_1$,
because
\be
\prod_{r=1}^{k-1} \left(1+n_1-2r \right)=0 \:\: \mbox{for } n_1 \mbox{ odd, }
1 \le n_1 \le k, \: k>1\: .
\ee

\section*{Acknowledgments} 

We would like to thank Dan Brace and Bogdan Morariu for many helpful
discussions.
This work was supported in part by the Director, Office of Science,
Office of High Energy and Nuclear Physics, Division of High Energy Physics of 
the U.S. Department of Energy under Contract DE-AC03-76SF00098 and 
in part by the
National Science Foundation under grant PHY-95-14797. B.L.C. is supported
by the DFG (Deutsche Forschungsgemeinschaft) under grant CE 50/1-1.


\begin{thebibliography}{99}

\bibitem{BMZ}
D.~Brace, B.~Morariu, B.~Zumino,
{\it Duality Invariant Born-Infeld Theory}, Yuri Golfand memorial volume; 
hep-th/9905218

\bibitem{ABMZ1}
P.~Aschieri, D.~Brace, B.~Morariu, B.~Zumino,
{\it Nonlinear Self-Duality in Even Dimensions}, Nucl.Phys. {\bf B574}:551-570,
2000 

\bibitem{ABMZ2}
P.~Aschieri, D.~Brace, B.~Morariu, B.~Zumino, {\it Proof of a symmetrized 
trace conjecture}, preprint LBNL-45360, LBL-45360, hep-th/0003228, March 2000,
to appear in Nucl. Phys. {\bf B}

\bibitem{Schwarz}
A.~Schwarz,
{\it Noncommutative Algebraic Equations and Noncommutative Eigenvalue 
Problem}, preprint hep-th/0004088, April 2000

\bibitem{Gant}
F.~R.~Gantmacher,
{\it The theory of matrices}, Chelsea Publishing Company, New York, 1977,
Vol I, pp. 81--83
\end{thebibliography}
\end{document}